\documentstyle[aps,prb,multicol,epsf,graphicx]{revtex}
\newcommand \bea {\begin{eqnarray}}
\newcommand \eea {\end{eqnarray}}
\newcommand \be {\begin{equation}}
\newcommand \ee {\end{equation}}

\newcommand \bi {\bibitem}

\begin{document}

\title{\bf\noindent The Nearest and 
Next Nearest Neighbor Sherrington-Kirkpatrick
Model}

\author{D.S. Dean$^{(1)}$ and F. Ritort$^{(2)}$}

\address{(1) IRSAMC, Laboratoire de Physique Quantique, Universit\'e Paul Sabatier, 118 route de Narbonne, 31062 Toulouse Cedex 04, France\\
(2) Departament de F\'{\i}sica Fonamental,
 Facultat de F\'{\i}sica, Universitat de Barcelona\\ Diagonal 647,
 08028 Barcelona, Spain\\
E-Mail:dean@irsamc.ups-tlse.fr, ritort@ffn.ub.es}

\maketitle
\begin{abstract}
The mean field theory of a spin glass with a specific form of nearest
and next nearest neighbor interactions is investigated. Depending
on the sign of the interaction matrix chosen we either find 
the  continuous replica symmetry breaking seen in the Sherrington-Kirkpartick
model or a one step solution similar to that found in structural glasses.
Our results are confirmed by numerical simulations and the link between 
the type of spin-glass behavior and the density of eigenvalues of
the interaction matrix is discussed.
\end{abstract}

%\begin{multicols}{2}
\vspace{.2cm}
\pagenumbering{arabic} 
\pagestyle{plain}

\section{Introduction}
The Sherrington-Kirkpatrick (SK) model \cite{SK} is the most
well-known example of a disordered and frustrated system in the field
of spin glasses \cite{REVIEW}. It corresponds to the infinite-range
version of the Edwards-Anderson (EA) model introduced earlier in 1975
\cite{EA}. In the EA model quenched disorder is introduced in the
random sign of the exchange couplings between nearest-neighbor spins
on a lattice. The infinite-ranged version is the natural mean-field
version of the EA model, in the same sense as the infinite-range
ferromagnet is the mean-field theory for the Ising model. A nearly
complete solution of the SK model has been found \cite{PARISI} which
has raised subtle questions about the nature of the spin-glass phase
showing that the mean field theory of spin glasses is considerably
more complex than the standard mean field of say ferromagnetic
systems.  Plenty of new questions were posed after it was shown that
the correct thermodynamic solution had to be understood in terms of a
replica-symmetric broken solution. For instance, does the ergodicity
breaking (implied by replica symmetry breaking) also occur in
short-range systems?  Also, has replica symmetry a true physical
meaning in as much as time-reversal symmetry has for usual
ferromagnets?  After many years of research these questions have
turned out to be extremely difficult and in the meantime criticisms
questioning their relevance regarding our understanding of
experimental systems have also been raised. Scaling theories of finite
dimensional spin glasses, so called droplet models \cite{DROP}, seem
to be at variance with the image of replica symmetry breaking.  It is
thus useful to introduce new solvable models which correspond to mean
field versions of different finite dimensional problems in order to
improve our understanding of the spin-glass problem.

In this paper we introduce a new solvable spin-glass model which
corresponds to the mean-field version of the Edwards-Anderson model
but including next-nearest neighbor interactions. We will refer to
this new model as NNN-SK model. The motivation is that the model
admits the possibility of being  realized on a finite-dimensional lattice
and incorporates correlations between the first nearest-neighbor and
second-nearest neighbor coupling interactions. At a first glance it seems
strange to consider a next nearest neighbor interaction in a totally
connected  spin-glass model, however we shall see that the correlations
introduced in the couplings can lead to a physics different to that
of the SK model. In addition this model allows one to discuss the role
of the density  eigenvalues of the interaction matrix in the spin-glass 
behavior. 

\section{The model}

The model we study is a totally connected one with Hamiltonian 
\begin{equation}
H = -{1\over 2} \sum_{ij} K_{ij} S_i S_j~~~~, \label{eqH}
\end{equation}
the spins $S_i$ ($1\leq i\leq N$) are Ising spins taking the values $\pm 1$. 
Here we take the interaction matrix $K$ to be a nearest neighbor and next
nearest neighbor type interaction

\begin{equation}
K = J J^T\label{KJ}
\end{equation}
where the $J_{ij}$ are independent Gaussian random variables
such that $\overline{J}_{ij} = 0$ and $\overline{J_{ij} J_{kl}} =
\delta_{ik}\delta_{jl}/N$. Here the over-line denotes the disorder
averaging and we note that the matrix $J$ here is not symmetric. 
The same model with $J$ symmetric may be studied, however taking $J$ 
non-symmetric considerably simplifies the analytical study of the model.

In this paper we will consider two models. The positive temperature
model where $K = J J^T$ (we will refer to this as the $K>0$ model) and the
negative temperature model where $K = -J J^T$ (we will refer to this
as the $K<0$ model). Contrarily to what happens in the usual
Sherrington-Kirkpatrick model, the model (\ref{eqH}) is not invariant
under the transformation $K\to -K$. Clearly $K$ is a matrix which is positive
definite and $-K$ a matrix which is negative definite.

For a symmetric matrix $J$ taken from the Gaussian ensemble, the density of
eigenvalues $\lambda$ is given by the Wigner semi-circle law
\cite{WIGNER}

\begin{equation} 
\rho_J(\lambda) = {1\over 2 \pi}\sqrt{4 - \lambda^2}
\end{equation}
with $\lambda \in [-2,2]$. In  corresponding the spherical spin model 
\cite{SKSPHER}, at low temperature, the system minimizes its energy 
via a macroscopic condensation onto the eigenvector corresponding
to $\lambda = 2$. In the case of the SK model this condensation is
not possible due to the discrete nature of the spins, however it seems
reasonable to assume that the density $\rho(\lambda)$ for 
$\lambda \sim 2$ plays an important role in the low temperature behavior.
In our model, if $J$ is taken to be symmetric, the density of eigenvalues
$\kappa$ is clearly  given by  
\begin{equation}
\rho_{\pm K}(\kappa) = {1\over 4 \pi\sqrt{\pm\kappa}}\sqrt{4 \mp \kappa}
\end{equation}
where now $\kappa \in [0,4]$ in the $+K$ or $K>0$ model and  $\kappa \in [-4,0]$ for the $-K$ or $K<0$ model.
In fact one can show that \cite{RHOCOM}  this is also 
the density of eigenvalues in the case of $J$ non-symmetric.
Hence in the positive case, the low energy region of the 
interaction matrix has the same form as the SK model and one
would have in a spherical model a condensation onto the eigenvector 
of largest eigenvalue. In the negative case however the largest
eigenvalues are at $\kappa = 0$, but the density of eigenvalues
now diverges in this region. In the corresponding spherical model
it is easy to see that this eliminates the finite temperature 
phase transition. This happens in exactly the same way that the 
divergence of the density of occupation of the  zero energy states for  
free bosons in  two or less  dimensions eliminates the finite temperature
Bose Einstein transition.

In the following treatment we consider the $K>0$ model although the same
mathematical treatment we present here can be applied to the $K<0$ case
on changing the sign of the inverse temperature $\beta$. 
The Hamiltonian of the model can thus be written as  

\begin{equation}
H = -{1\over 2} \sum_{ij} J_{ik} J_{jk} S_i S_j
\end{equation}
and  the partition function for the model is therefore given by
\begin{equation}
Z = {\rm Tr}_{S_i} \exp\left[ {\beta \over 2}\sum_{ij} J_{ik} J_{jk} S_i S_j \right]
\end{equation}
In order to facilitate taking the disorder average we make a Hubbard 
Stratonovich transformation by introducing the auxiliary Gaussian spins
$x_i$ to obtain
\begin{equation}
Z = {\rm Tr}_{S_i, x_i} \exp\left[ {\beta \over 2}\sum_{ij} J_{ij} S_i x_j \right] \label{eqlittle}
\end{equation}
where ${\rm Tr}_{x}$ indicates the trace over the Gaussian spin $x$ and is 
defined by 
\begin{equation}
{\rm Tr}_{x} = \int dx\  \sqrt{{\beta \over 2 \pi}}\exp(-{\beta x^2\over 2})
\end{equation}
In the form (\ref{eqlittle}) the model is that of an asymmetric Little
model \cite{LITTLE} but where one of the two sets of spins is
Gaussian  rather than Ising. Interest in the Little model arose one decade
ago, in the context of neural networks as the parallel dynamics of
the standard Hopfield model \cite{HOP} coincides with the sequential 
dynamics in the Little model \cite{CMPP}. 
The corresponding mean-field
spin-glass model  was studied in \cite{BPR} where it was shown that the 
equilibrium behavior is the same for both the  SK and Little (for both 
asymmetric and symmetric couplings)
models.
Here we have verified numerically that, in our model also, a symmetric or 
asymmetric matrix leads to the same physical behavior.

\section{The Positive Temperature $K>0$ Model}

We introduce replicas of both the Gaussian and Ising spins in order
to average over the disorder via the replica method,  obtaining
\begin{eqnarray}
\overline{Z^n} &=& {\rm Tr}_{S_i^a, x^a_i} \exp\left[ {\beta^2\over 2 N} 
\sum_{ij} \sum_{ab} S_i^a x_j^a S_i^b x_j^b \right] \nonumber \\
&=& {\rm Tr}_{S_i^a, x^a_i} \exp\left[ {\beta^2 N\over 2}\sum_{ab} Q_{ab}P_{ab}
\right]
\end{eqnarray}

where we have introduced the Ising and parabolic spin overlaps

\begin{equation} 
Q_{ab} = {1\over N} \sum_{i} S_i^a S_i^b \ \ \ {\rm and} \ \ \
P_{ab} = {1\over N} \sum_{i} x_i^a x_i^b
\end{equation}
The trace over the Ising spins is accomplished using a delta function
representation of the overlap constraint

\begin{eqnarray}
{\rm Tr}_{S_i^a}
\delta\left( {N\over 2} Q_{ab}  - \sum_{i} S_i^a S_i^b \right) &=&
{1\over (2\pi)^{n^2}}
{\rm Tr}_{S_i^a}\int d\Lambda_{ab} \exp\left( {N\over 2} \sum_{ab} 
\Lambda_{ab} Q_{ab}
- {1\over 2} \Lambda_{ab} \sum_i S_i^a S_i^b \right) \nonumber \\
&=& {1\over (2\pi)^{n^2}}
\int d\Lambda_{ab} \exp\left( {N\over 2} \sum_{ab} \Lambda_{ab} Q_{ab}
+ N \ln(Z_S)\right)
\end{eqnarray}
where 
\begin{equation}
Z_S = {\rm Tr}_{S^a} \exp\left(-{1\over 2} \sum_{ab} \Lambda_{ab} S^a S^b\right)
\end{equation}
The same procedure is used for the Gaussian spins to yield
\begin{equation}
{\ln\left[ \overline{Z^n} \right]\over N}
= \rm{extr}\   S^{**}[Q,P,\Lambda, \Gamma]
\end{equation}
where $S^{**}[Q,P,\Lambda, \Gamma]$ is the saddle point action 
\begin{eqnarray}
 S^{**}[Q,P,\Lambda, \Gamma]
&=& {\beta^2\over 2}\sum_{ab} Q_{ab} P_{ab} + {1\over 2}\sum_{ab} \Lambda_{ab}
Q_{ab} +{1\over 2}\sum_{ab} \Gamma_{ab}
P_{ab}  \nonumber \\
&+& \ln(Z_S) + \ln(Z_x) \label{eqsp1}
\end{eqnarray}
and 
\begin{equation}
Z_x = {\rm Tr}_{x^a} \exp\left( -{1\over 2} \sum_{ab} \Gamma_{ab} x^a x^b 
\right)
\end{equation}
The saddle point equations with respect to $P_{ab}$ and $Q_{ab}$ allows
the evaluation of the saddle values   of the variables 
$\Lambda_{ab}$ and $\Gamma_{ab}$
\begin{eqnarray}
{\partial S^{**} \over \partial P_{ab}} &=& 0 \ \Rightarrow \ \Gamma_{ab} = 
- \beta^2 Q_{ab}
\\
{\partial S^{**} \over \partial Q_{ab}} &=& 0 \ \Rightarrow \ \Lambda_{ab} = - \beta^2 P_{ab}
\end{eqnarray}
leading to the reduced saddle point  $S^{*}[Q,P]$ such that
\begin{equation}
{\ln\left[ \overline{Z^n} \right]\over N}
= \rm{extr} \   S^{*}[Q,P]
\end{equation}
with
\begin{equation}
S^{*}[Q,P] = -{\beta^2\over 2}\sum_{ab} Q_{ab} P_{ab}
+ \ln\left[ {\rm Tr}_{S^a} \exp({\beta^2\over 2}\sum_{ab} P_{ab} S^a S^b)\right]
-{1\over 2} Tr\log\left( I - \beta Q\right)
\end{equation}
and where the Gaussian integral giving $Z_x$ has been evaluated.
The saddle point value of $P_{ab}$  is given by
\begin{equation}
{\partial S^* \over \partial Q_{ab}}\ \Rightarrow \ P = {1\over \beta}(I - \beta Q)^{-1}
\end{equation}
thus leading to the new reduced action $S[Q]$ such that
\begin{equation}
{\ln\left[ \overline{Z^n} \right]\over N}
= \rm{extr} \  S[Q]
\end{equation}
with
\begin{equation}
S[Q] = {1\over 2}\left( n - Tr\ (I - \beta Q)^{-1} - Tr \ln(I - \beta Q) \right) + \ln {\rm Tr}_{S^a}  \exp\left({\beta\over 2}\ \sum_{ab}\left[(I - \beta Q)^{-1}\right]_{ab} S^a S^b \right) \label{eqsqab}
\end{equation}
The saddle point equation from (\ref{eqsqab}) is 
\begin{equation} 
-{\beta \over 2}(I-\beta Q)^{-2} + {\beta\over 2} (I-\beta Q)^{-1}
+{\beta^2\over 2}(I-\beta Q)^{-2} A = 0
\end{equation}
where the elements of the matrix $A$ are given by
\begin{equation}
A_{ab} = { {\rm Tr}_{S^a} S^a S^b \exp\left({\beta\over 2}\ \sum_{ab}\left[(I - \beta Q)^{-1}\right]_{ab} S^a S^b \right) \over 
{\rm Tr}_{S^a}  \exp\left({\beta\over 2}\ \sum_{ab}\left[(I - \beta Q)^{-1}\right]_{ab} S^a S^b \right)}
\end{equation}
We use the fact that the matrix $(I-\beta Q)$ should not be singular at the 
saddle point (otherwise the saddle point lies on a branch cut) to conclude
\begin{equation}
Q = A \label{eqphysq}
\end{equation}
which is the physical saddle point equation for $Q$.
We proceed by studying the replica symmetric from of this
action with

\begin{equation}
Q^{RS}_{ab} = (1-q) \delta_{ab} + q U_{ab}
\end{equation}
where $U_{ab} = 1$ for all $(a,b)$. After some algebra one finds that

\begin{eqnarray} 
s(q) &=& \lim_{n\to 0}{S[Q^{RS}]\over n} = -{1\over 2}
\ln\left( 1 - \beta(1-q)\right) - {1\over 2} {\beta q (-1 + 2 \beta q (1-q))
\over (1 - \beta(1-q))^2} \nonumber \\
&+&\int {dz\over \sqrt{2\pi}} \ \exp\left(-{z^2\over 2}\right) \ln\left( 2 \cosh\left( {\beta z 
\sqrt{q} \over (1 - \beta (1-q))}\right)\right) \label{eqsrs}
\end{eqnarray}
Taking the saddle point with respect to $q$ above yields
\begin{equation}
{ds(q)\over dq} = {\beta ^2 \over 2}{(\beta q + \beta -1)\over 
(1 - \beta(1-q))^3 }\left(\int {dz\over \sqrt{2\pi}} \ \exp\left(-{z^2\over 2}\right) \tanh^2\left(( {\beta z 
\sqrt{q} \over (1 - \beta (1-q))}\right) -q \right) = 0 \label{eqq0f}
\end{equation}
There are two sets of possible solutions to Eq. (\ref{eqq0f})
\begin{equation}
q = {1\over \beta} -1 \label{eqqa}
\end{equation}
and 
\begin{equation}
q = \int {dz\over \sqrt{2 \pi}} \ \exp\left(-{z^2\over 2}\right) \tanh^2\left(( {\beta z 
\sqrt{q} \over (1 - \beta (1-q))}\right) \label{eqqb}
\end{equation}
However as mentioned above the solution (\ref{eqqa}) is unphysical.
The solution (\ref{eqqb}) corresponds to the physical solution
(\ref{eqphysq}) and only has the paramagnetic 
solution $q=0$ for $T>T_c=2$. Hence for $T>T_c=2$ one is (in the replica symmetric
scheme) in a paramagnetic phase. 

The energy per spin $E$  obtained from (\ref{eqsqab}) is
\begin{equation}
E = \lim_{n\to 0}{1\over 2\beta}\left[ 1 - {1\over n} 
{\rm Tr} (1 -\beta Q)^{-1}\right]
\end{equation} 
where $Q_{ab}$ here is the solution to the saddle point equation.
The replica symmetric energy per spin is thus
\begin{equation}
E^{RS} = -{1\over 2}\left( {(1-q)(1- \beta (1-q)) + q\over (1- \beta (1-q))^2}
\right)
\end{equation}
Hence for  $T>T_c=2$, that is to say in the paramagnetic phase where one
can be sure the the RS solution is correct, one has
\begin{equation}
E = - {1\over 2(1 - \beta)}
\end{equation}
For $T \ll 1$ a low temperature expansion of Eq. (\ref{eqqb}) yields
\begin{equation} 
q = 1 - {T\over 1 + \sqrt{{\pi\over 2}}} + O(T^3)
\end{equation}
With the replica symmetric ansatz, in the spin-glass phase,
the ground state energy is
\begin{equation}
E^{RS}_{GS} = -{1\over 2}( {2\over \pi} + {2 \sqrt{2}\over \sqrt{\pi}}
+1)
\end{equation}
As in the SK model if one computes the entropy corresponding to this
replica symmetric solution within the spin-glass phase one finds a 
temperature where it becomes negative, indicating the need to 
break replica symmetry.
The continuous nature of the transition however suggests that the 
underlying physics is the same as that of the SK model and the Landau 
expansion of $S[Q_{ab}]$ has the same generic form as that of the $SK$ model
indicating a continuous RSB which sets in at $T_c$. 
The introduction of NNN interactions thus does not qualitatively change
the behavior of the model, in agreement with the previously discussed
intuitive notion that (for a given spin type) the 
behavior of the density of the largest eigenvalues determines the 
nature of the transition.  
\section{The Negative Temperature $K<0$ Model}
The Hamiltonian (\ref{eqH}) is not  invariant under the transformation
$K \to -K$. In this case the action for the replicated partition function is
given by

\begin{equation}
S[Q] = {1\over 2}\left( n - Tr\ (I + \beta Q)^{-1} - Tr \ln(I +\beta Q) \right) + \ln {\rm Tr}_{S^a}  \exp\left(-{\beta\over 2}\ \sum_{ab}\left[(I + \beta Q)^{-1}\right]_{ab} S^a S^b \right) \label{eqsqab2}
\end{equation}

In the high temperature phase we expect the annealed approximation 
to be exact and find

\begin{equation}
f_{\rm ann} = -{1\over \beta}\ln(2) + {1\over 2\beta } \ln(1 + \beta)
\end{equation}

The annealed entropy per spin $S_{\rm ann}$ is thus

\begin{equation}
S_{\rm ann} = -\beta^2 {\partial f \over \partial \beta}
= \ln(2) - {1\over 2} \ln( 1+ \beta) + {\beta \over 2  (1+ \beta)}
\end{equation}

The annealed energy is given by

\begin{equation}
E_{\rm ann} = {1\over 2 ( 1+ \beta)}
\end{equation}

The annealed solution can only be valid as long as $S_{\rm ann}$ is
positive.  If there exists a $\beta_{\rm bound}$ such that $S_{\rm
ann}(\beta_{\rm bound}) < 0$ for $\beta > \beta_{\rm bound}$, then
$\beta_{\rm bound}$ gives us an upper bound for the inverse
temperature at which the equilibrium transition occurs, $\beta_K$.  In
the Adam-Gibbs-Di Marzio scenario \cite{AGM} $T_K=1/\beta_K$ corresponds to the
temperature where the configurational entropy $S_{\rm conf}$ vanishes
and the subindex $K$ in $\beta$ stands for the Kauzmann temperature which
originates from the Kauzmann paradox \cite{K} which is based on the observation
that an extrapolation of a high-temperature entropy cannot cross the
low temperature solution. Sometimes $T_K$ is also referred to as the 
ideal glass transition temperature not to be confused with the
experimental glass transition temperature $T_g$ in the context of finite
dimensional glasses which turns out to be only a convention
corresponding to an extremely large relaxation time as opposed to 
a diverging one.  Because both
entropies ($S_{\rm ann}$ and $S_{\rm conf}$) are monotonically
increasing functions of the temperature and $S_{\rm conf}<S_{\rm ann}$
this means that $\beta_{\rm bound}>\beta_K$ is an upper bound of the
ideal glass transition temperature $T_K$.  One finds that $\beta_{\rm
bound} \approx 8.82$ and hence $T_{\rm bound} = 1/\beta_{\rm bound}\approx  0.113$.

In the replica symmetric ansatz the equation for the Edwards-Anderson
order parameter $q$ is

\begin{equation}
q = \int {dz\over \sqrt{2\pi}}\exp\left(-{z^2\over 2}\right) \tanh^2\left(( {\beta z 
\sqrt{q} \over (1 + \beta (1-q))}\right) \label{eqrsf}
\end{equation}

The examination of this equation shows that one can no longer have a continuous
phase transition  from $q = 0$. For low temperatures one can show that
a solution with $q$ non zero exists and that

\begin{equation}
q = 1 - {T\over \sqrt{{\pi\over 2}} -1} + O(T^2)
\end{equation}

Numerically the inverse temperature $\beta_c^{RS}$ at which $q$ can become
non zero is found to be  $\beta_c^{RS} \approx 29.3$, thus giving
$T_c^{RS}\approx 0.034$. This is much lower than $T_K$ above and hence
the replica symmetric solution cannot eliminate the entropy crisis at
$T_{\rm bound}$. Clearly one must resort to a replica symmetry broken ansatz. 
Guided by the results of our RS calculation and the numerical simulations
we make the REM (one step) like ansatz where the matrix $Q$ is given by $n/m$ 
matrices $\tilde{Q}$ about the diagonal of  $Q$ of size $m\times m$ and is zero
outside these blocks. The matrix $\tilde{Q}$ takes the form 
\begin{equation}
\tilde{Q} = I (1-q) + q {\tilde U}
\end{equation}
where $I$ is the identity matrix and  ${\tilde U}$ is the 
matrix with each element equal to  $1$.
We note the following results
\begin{equation}
(I + \beta \tilde{Q})^{-1} = {1\over 1 +\beta(1-q)}I 
+ {\beta q{\tilde U} \over (1+\beta(1-q))(1+\beta(1-q) + \beta mq)}
\end{equation}
\begin{eqnarray}
{\rm Tr} (1 +\beta Q)^{-1} &=& {n\over m} {\rm Tr} (1 +\beta {\tilde Q})^{-1}
\nonumber \\
&=& {n\over m}\left( {1\over 1 +\beta (1-q) + \beta mq} + {m-1 \over
1 + \beta(1-q)}\right)
\end{eqnarray}
\begin{eqnarray}
{\rm Tr} \ln\left[(1 +\beta Q)^{-1}\right] &=& {n\over m} {\rm Tr} \ln\left[(1 +\beta {\tilde Q})^{-1}\right]
\nonumber \\
&=& {n\over m}\left( \ln( 1 + \beta(1-q) + \beta q m) + (m-1) \ln(1 +
\beta(1-q))\right)
\end{eqnarray}
and 
\begin{eqnarray}
\ln {\rm Tr}_{S^a}  \exp\left({\beta\over 2}\ \sum_{ab}\left[(I - \beta Q)^{-1}\right]_{ab} S^a S^b \right) &=& 
{n\over m} \ln {\rm Tr}_{S^a}  \exp\left({\beta\over 2}\ \sum_{ab}\left[(I + \beta {\tilde Q})^{-1}\right]_{ab} S^a S^b \right)
\nonumber \\
&=& {n \beta \over 2 (1 + \beta(1-q))} \nonumber \\
&+& {n\over m}
\ln\left[ \int {dz\over \sqrt{2 \pi}} 
\exp(-{1\over 2} z^2)\left(2 \cosh(\beta z \alpha(q,m,\beta))\right)^m
\right]
\end{eqnarray}
where 
\begin{equation}
\alpha(q,m,\beta) = {\sqrt{q}\over \sqrt{(1 +\beta (1-q))(1 + \beta(1-q) + \beta mq)}}
\end{equation}
One therefore finds the one step action
\begin{eqnarray} 
s(q,m) &=& \lim_{n\to 0}{S[Q^{1RSB}]\over n} \nonumber \\
&=& {1\over 2}\left[ {-\beta^2 q ( 1-q + mq)\over (1+ \beta(1-q))
(1 + \beta(1-q) + m\beta q)} -{1\over m}\ln\left(1 + \beta (1-q) + m\beta q
\right)\right.
\nonumber \\
&-& \left. (1-{1\over m}) \ln\left(1 + \beta(1-q)\right)\right]
+ {1\over m}\ln\left( z(q,m)\right)\label{sqm}
\end{eqnarray}
where
\begin{equation}
z(q,m) = \int {dz \over \sqrt{2\pi}} \exp(-{1\over 2} z^2)
\left(2 \cosh(\beta z \alpha(q,m,\beta))\right)^m
\end{equation}
One finds that
\begin{eqnarray}
{\partial\over \partial q} s(q,m) &=&
\beta^2 \alpha(q,m,\beta) {\partial \alpha(q,m,\beta) \over \partial q}(1-m)
\nonumber \\
&\times& \left[ q - {1\over z(q,m)}\int {dz \over \sqrt{2\pi}} \exp(-{1\over 2} z^2)
\left(2 \cosh(\beta z \alpha(q,m,\beta))\right)^m \tanh^2(\beta \alpha(q,m,\beta) z)\right]
\end{eqnarray}
The saddle point of $s(q,m)$  corresponding to the solution (\ref{eqphysq}) is
thus
\begin{equation}
q = {1\over z(q,m)}\int {dz \over \sqrt{2\pi}} \exp(-{1\over 2} z^2)
\left(2 \cosh(\beta z \alpha(q,m,\beta))\right)^m \tanh^2(\beta \alpha(q,m,\beta) z)
\label{eqqm}
\end{equation}
If one considers the case $m=1$ this solution must give the same
free-energy as the annealed free energy. However one can find a
non zero value of $q$ which signals a dynamical transition where
the system becomes stuck in metastable states of high free energy.
Setting $m=1$ in Eq. (\ref{eqqm}) gives
\begin{equation}
q = \exp\left(-{1\over 2}\beta^2\alpha(q,1,\beta)^2\right)
\int {dz \over \sqrt{2\pi}} \exp(-{1\over 2} z^2)
\cosh(\beta z \alpha(q,1,\beta)) \tanh^2(\beta \alpha(q,1,\beta) z)
\label{eqqma}
\end{equation}
For small $\beta$ this equation has only the solution $q= 0$. However
at $\beta_d \approx 7.325$ ($T_d = 1/\beta_d \approx 0.137$) one finds
a non zero value of $q$ with $q \approx 0.922$. This transition
corresponds to what is known as the mode-coupling transition in
mode-coupling theories of the glass transition in their idealized
version \cite{MCT}.

The precise way to locate both transitions ($T_d$ and $T_K$) was
suggested in a series of papers by Kirkpatrick, Thirumalai and Wolynes
and collaborators \cite{KTW} and later on applied to several models
such as the random orthogonal model \cite{ROM}, Potts glasses
\cite{POTTS} and mean-field quantum spin glasses \cite{Q}. The static
transition is located by expanding the corresponding free energy
$f(q,m)$ about $m = 1$ and writing

\begin{equation} 
\beta f(q,m) = \beta f_{\rm para} +  (1-m) V(q) + O\left( (1-m)^2\right)
\end{equation}
where $f_{\rm para} = f(0,1)$ is the paramagnetic free energy and
\begin{equation}
V(q) = -\beta \frac{\partial f(q,m)}{\partial m}\vert_{m=1} =  
\frac{\partial s(q,m)}{\partial m}\vert_{m=1}
\end{equation}
The static transition $T_K$ is given by the values of $\beta_K$ where
$V(q_K) = V'(q_K) = 0$. The dynamic transition $T_d$ is given by the
conditions $V'(q_d) = V''(q_d) = 0$ which correspond to the marginality
condition and coincide with the solution found at $m=1$ and reported above.

\begin{figure}[tbp]
\begin{center}
\rotatebox{0}
{\includegraphics*[width=9cm,height=9cm]{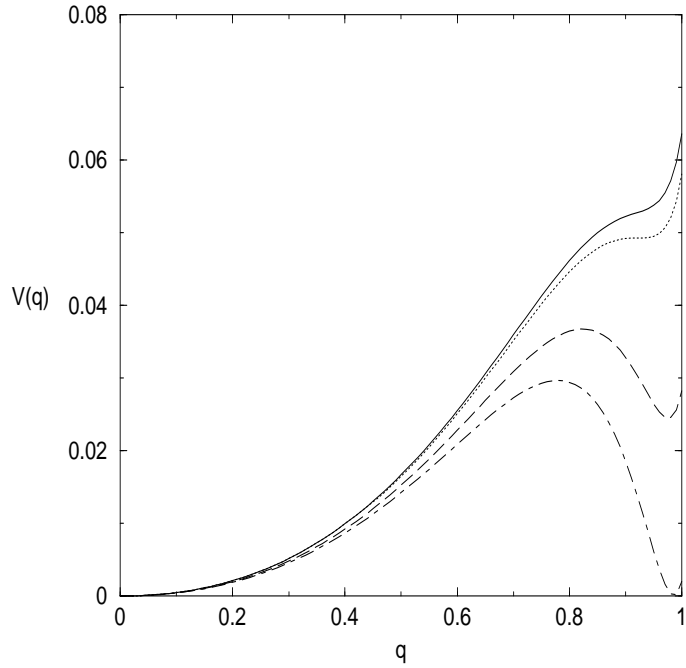}}
\caption{Behavior of the potential $V(q)$ on lowering $T$. The different
regimes are: $T > T_d$ (solid line), $T = T_d$ (dotted line), $T_K < T < T_d$
(dashed line) and $T = T_K$ (dotted dashed line).}\label{figvq}
\end{center}
\end{figure}

The explicit form of the potential $V$ is 
\begin{eqnarray}
-V(q) &=& {\beta q ( 1 + 2\beta)\over 2(1+\beta) (1 + b (1-q))}
+ {1\over 2} \ln( 1 +\beta (1-q)) 
- {1\over 2} \ln(1 + \beta) \nonumber \\
&+& \exp\left(-{1\over 2}\beta^2\alpha(q,m,\beta)^2\right)\int {dz \over \sqrt{2\pi}} \exp(-{1\over 2} z^2)
\cosh(\beta z \alpha(q,1,\beta)) \ln\left(\cosh(\beta \alpha(q,1,\beta) z)\right) \nonumber \\
\end{eqnarray}
Using this to compute  the static transition one finds that
this gives $T_K \approx 0.116$
and  $q_K \approx 0.985$. The behavior of $V(q)$ on lowering $T$ is shown in 
Fig. (\ref{figvq}).

Note that at $T_d$ the value of the potential $V(q_d)$ is finite and
corresponds to the configurational entropy at that temperature. As $T$
decreases below $T_d$, the analytical continuation of the solution
$q_d$ to the saddle point of  Eq.(\ref{sqm}) increases while the
value of $m_d(T)$ decreases. The configurational entropy evaluated at
$T<T_d$ is $S_c(T)=\beta(f(q_d(T),m_d(T))-f_{\rm para}(T))$ and
decreases while T decreases down to a temperature $T_K$ where it
vanishes and $q_K$ is the analytically continued value of $q_d(T_K)$.

Hence we see that in the case $K < 0$ the physics of the problem is
drastically altered by the interaction matrix, and that the model now
exhibits the phenomenology of a (mean field) structural glass.

\section{Numerical simulations.}

We have verified the main predictions for the NNN-SK model with positive
and negative temperature by doing some numerical simulations in both
cases. For the positive temperature model we checked that the transition
indeed occurs at $T=2$. This has been done  applying standard finite-size scaling
techniques useful for  investigating small-size systems. For the negative
model we found that strong freezing occurs at the mode-coupling
temperature $T_d=0.137$ as happens for other models such as the
$p$-spin \cite{CHS} or
the random orthogonal model \cite{ROM}.

\subsection{Some details of the simulations}

Simulations consist of standard Monte Carlo annealings using the
Metropolis algorithm. The system is cooled down from high temperatures
(typically twice the value of $T_c$ for the $K>0$ model and twice
$T_d$ for the $K<0$ case). Annealing schedules are as follows: every
$\Delta T=0.2$ for the $K>0$ model and every $\Delta T=0.01$ for the
$K<0$ model the system is allowed to equilibrate over $1000$ Monte Carlo
steps (MCs) and statistics are collected during $10^5$ MCs at each
MCs. The sizes are small, $N=25,50,75,100$, but enough to
locate the transition with some precision.  The simulated range of
temperatures are from $T=4$ down to $T=0.2$ for the $K>0$ model and from
$T=0.3$ down to $T=0.01$ for the $K<0$ model. The number of samples where 
several thousands for all sizes.

Due to the long-range character of the interactions the dependence of
the time needed to do a MCs grows quite fast with the size of the system
(actually like $N^2$). Therefore, for the statics, we had to limit our
investigation to relatively small sizes. Moreover, a careful study of
the relevant parameters for the transition (such as the kurtosis or the G
parameter to be defined below) requires a large number of samples (this
is especially true for parameters like G which measure sample-to-sample
fluctuations). This last parameter is the most successful example of what
are referred to as order-parameter fluctuation parameters (OPF
parameters) \cite{OPF}.

\begin{figure}[tbp]
\begin{center}
\rotatebox{270}
{\includegraphics*[width=7cm,height=10cm]{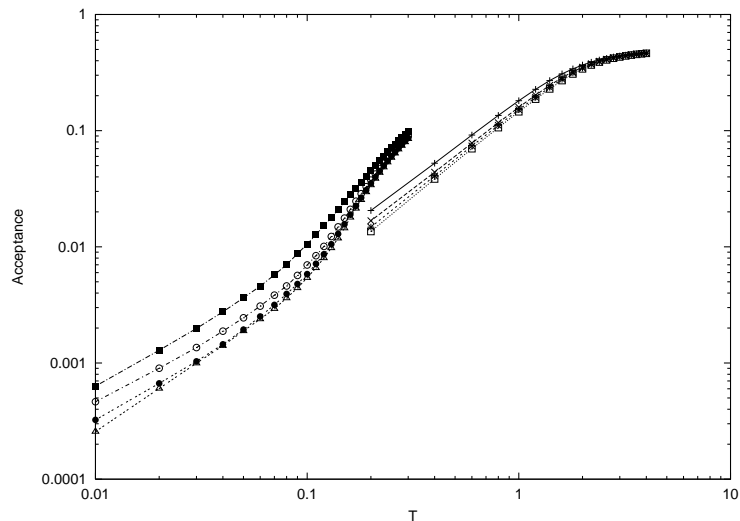}}
\caption{Acceptance rate as a function of $T$ in the relevant temperature
ranges for the $K>0$ (right set of curves) and $K<0$ (left set of
curves) models. For each set and from top to bottom we have
$N=25,50,75,100$. Note that the acceptance rate is typically 10 times
smaller in the $K<0$ model.\label{accep}}
\end{center}
\end{figure}

Before showing the results let us mention that, while for the positive
K model we achieved thermalisation in a range of temperatures in the
vicinity of $T_c$, for the negative $K$ model thermalisation was hardly
achieved due to the quite small acceptance rate for all the temperatures
simulated. This behavior is due to the small value of the relevant
temperatures of the $K<0$ model where the transition occurs (one order
of magnitude smaller). Because the typical energy change for both
models is the same, the Boltzmann factor is drastically reduced for
changes which increase the energy in the $K<0$ model as compared to the
$K>0$ case. This implies a very small acceptance rate for the $K<0$
models as compared to the $K>0$ case. In Fig.~(\ref{accep}) we show
the acceptance as a function of $T$ for the two cases. Note that the
acceptance is nearly two orders of magnitude smaller in the negative
model as compared to the positive model. Hence a good sampling of the
configurational space for the negative model can be excluded.

\subsection{The positive temperature $K>0$ model}

\begin{figure}[tbp]
\begin{center}
\rotatebox{270}
{\includegraphics*[width=7cm,height=10cm]{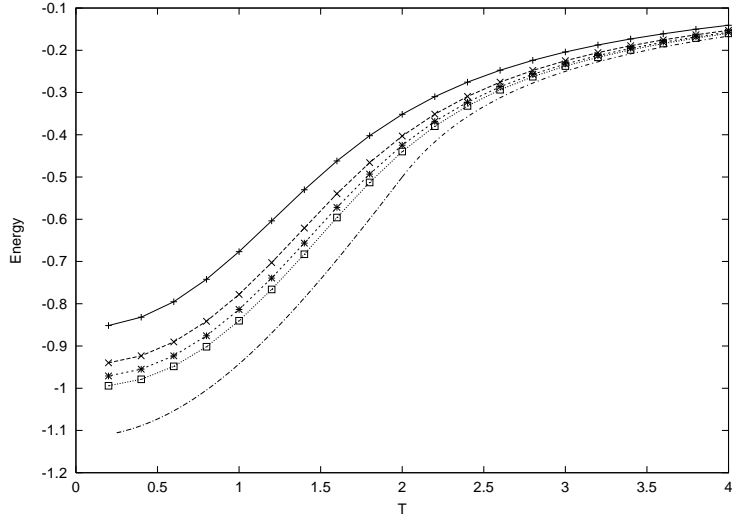}}
\caption{Energy as a function of $T$ for the model $K>0$. From top to
bottom we have $N=25,50,75,100$. The solid line is the replica
symmetric result.}
\label{ene+}\end{center}
\end{figure}

\begin{figure}[tbp]
\begin{center}
\rotatebox{270}
{\includegraphics*[width=7cm,height=10cm]{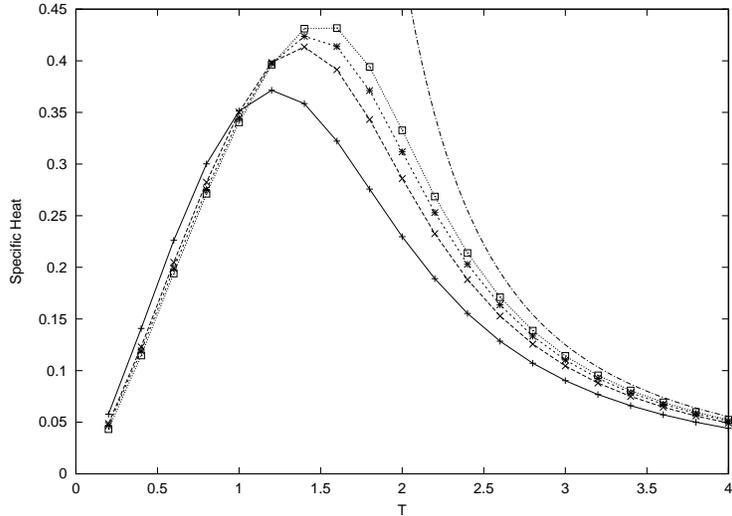}}
\caption{Specific heat as a function of $T$ for the model $K>0$. From
bottom to top at high T we have $N=25,50,75,100$.The solid line is
the high-T result.}
\label{c+}\end{center}
\end{figure}

In this model there is a continuous RSB transition at $T_c=2$. The
energy and the specific heat are shown as a function of $T$ in
Figs.~(\ref{ene+}) and (\ref{c+}). We also plot the result for
the annealed expression only valid above the critical temperature. As we
see the behavior of these quantities is similar to what is found for
the SK model: the maximum of the specific heat occurs below $T_c$.

The transition can be well determined by looking at the B and G
parameters as functions of the temperature. These parameters are defined by,

\begin{eqnarray} B=\frac{1}{2}\Bigl(3-\frac{\overline{\langle
q^4\rangle}}{\bigl(\overline{\langle q^2\rangle}\bigr)^2}\Bigr)
\label{eqB}\\
 G=\frac{\overline{\bigl(\langle
q^2\rangle\bigr)^2}-\bigl(\overline{\langle
q^2\rangle}\bigr)^2}{\overline{\langle
q^4\rangle}-\bigl(\overline{\langle q^2\rangle}\bigr)^2}
\label{eqG}
\end{eqnarray}

In the infinite-size limit these parameters behave as:
$B(T)=\hat{B}(T) \theta_H(T_c-T)$ while the behavior of $G$ turns out
to be simpler, $G(T)=\frac{1}{3} \theta_H(T_c-T)$   and hence transpires to be  a
better indicator for the transition. In Fig.~(\ref{eqB}) we show the
kurtosis parameter $B$ as a function of $T$ for different
sizes. In the same way as in the SK model \cite{OPF} we find a crossing at a
temperature close to $T_c=2$. This crossing turns out to be also present
for the OPF parameter $G$

\begin{figure}[tbp]
\begin{center}
\rotatebox{270}
{\includegraphics*[width=7cm,height=10cm]{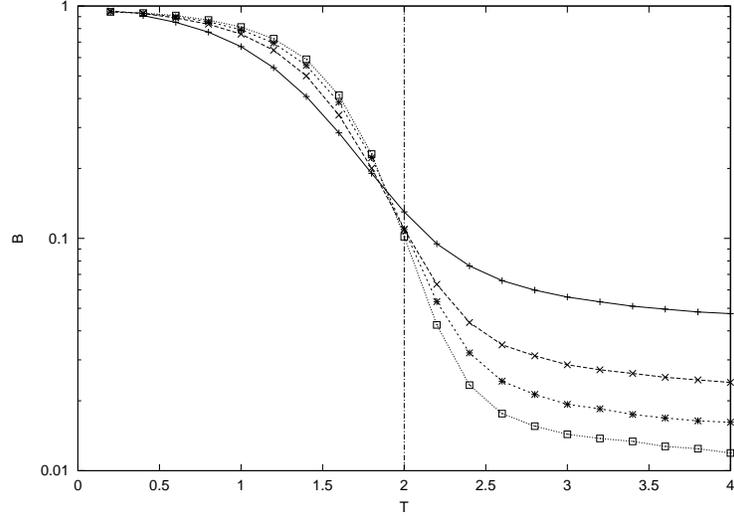}}
\caption{Kurtosis B as a function of $T$ for the model $K>0$. From top to
bottom we have $N=25,50,75,100$. The crossing temperature shifts with
the size towards $T_c=2$.}
\label{B+}\end{center}
\end{figure}

\begin{figure}[tbp]
\begin{center}
\rotatebox{270}
{\includegraphics*[width=7cm,height=10cm]{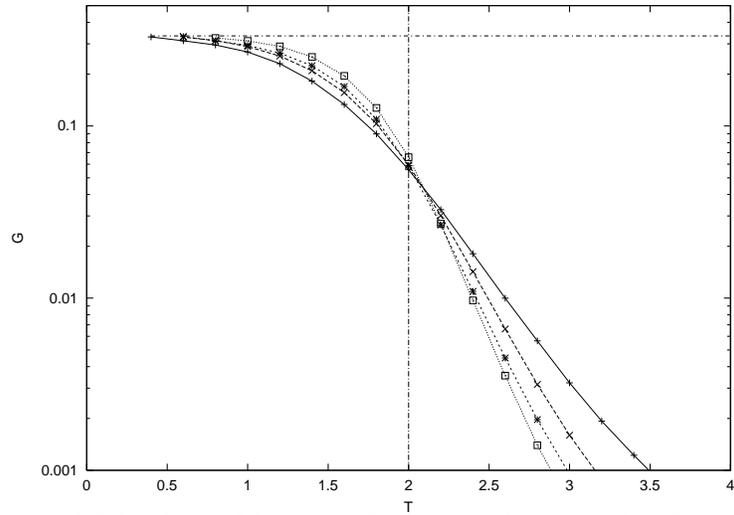}}
\caption{OPF G as a function of $T$ for the model $K>0$. From top to
bottom at low T we have $N=25,50,75,100$. The different curves cross
close to $T_c=2$ where $G_c\simeq 0.057$.}
\label{G+}\end{center}
\end{figure}

\subsection{The negative temperature $K<0$ model}

\begin{figure}[tbp]
\begin{center}
\rotatebox{270}
{\includegraphics*[width=7cm,height=10cm]{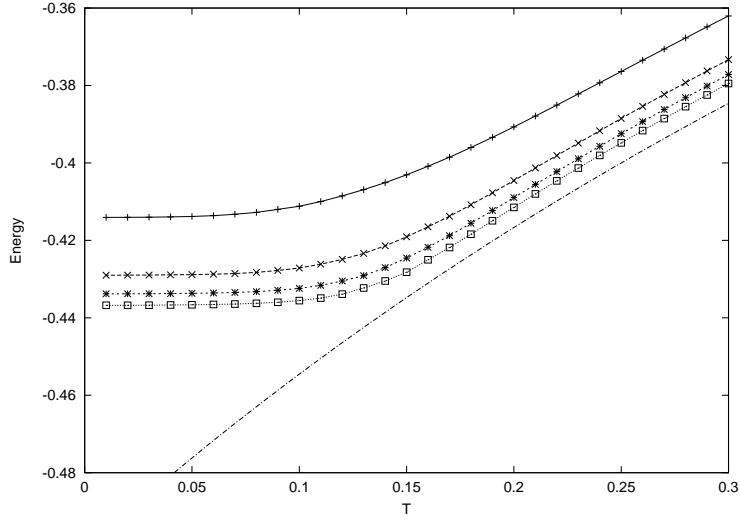}}
\caption{Energy as a function of $T$ for the model $K<0$. From top to
bottom we have $N=25,50,75,100$. The solid line is the high-T
result.}
\label{ene-}\end{center}
\end{figure}

\begin{figure}[tbp]
\begin{center}
\rotatebox{270}
{\includegraphics*[width=7cm,height=10cm]{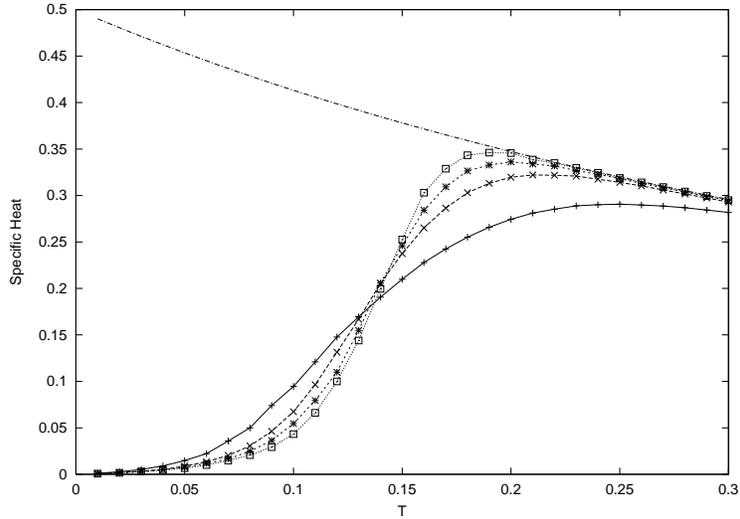}}
\caption{Specific heat as function of $T$ for the model $K<0$. From top to bottom at
low $T$ we have $N=25,50,75,100$.The solid line is the high-T
result.}
\label{c-}\end{center}
\end{figure}

As mentioned previously, we were not able to thermalise for this case close to
the transition so we do not have good data for order parameters such as B
or G. We show the results for the energy and specific heat in 
Figs.~(\ref{ene-},\ref{c-}) .

As we said before we did not succeed in thermalising the model at low
temperatures to extract the behavior of the order
parameters. Consequently we do not have good data for the kurtosis $B$
(\ref{eqB}) and the OPF $G$ (\ref{eqG}). The only transition about which we
may have some hints is the dynamical transition where dynamics becomes
extremely slow. One could be tempted to interpret the crossing point for
the specific heat at different sizes (Fig. (\ref{c-})) as the
signature of that dynamical transition. Despite of the fact that this
interpretation seems reasonable we are not absolutely certain and 
so we prefer to leave this  question open.

\begin{figure}[tbp]
\begin{center}
\rotatebox{270}
{\includegraphics*[width=7cm,height=10cm]{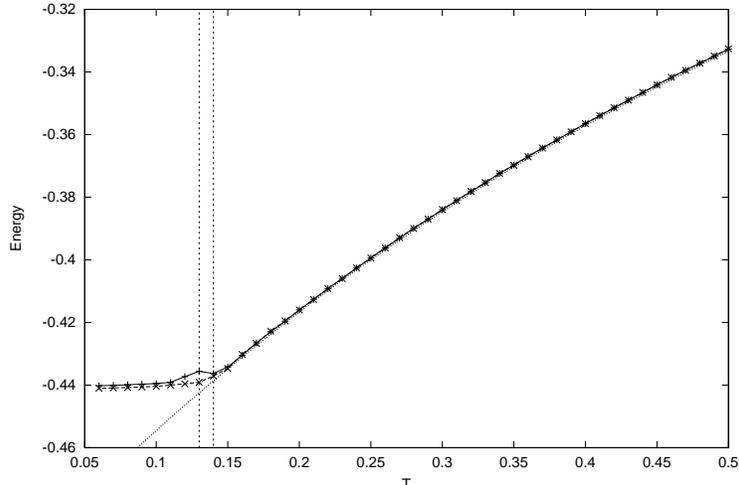}}
\caption{Cooling experiments in the model $K<0$. The two vertical dashed
lines limit the range of temperatures $0.13-0.14$. The dotted
line is the high-T result}
\label{cooling}\end{center}
\end{figure}

More evidence for the dynamical transition can be obtained by doing
annealing experiments for finite cooling rates and very large sizes. One
of these cooling experiments is shown in Fig.~(\ref{cooling}) for  a
system of size $N=1000$ was cooled down to very low temperatures 
staying for $10^5$ MCs at each
temperature and changing $T$ by $0.01$. The system gets trapped below a
crossover temperature $T^*\sim 0.14$ and starts to deviate from the
high-T line. We plot two independent dynamical histories to demonstrate
how the departure from the ergodic line depends on the dynamical history
signaling the presence of non-ergodic effects in the dynamics (below
$T^*$ the typical relaxation time has become much larger than the
time the system remains at each temperature, i.e. $10^5$).

Note that in this dynamical study the static transition $T_K$ remains
completely hidden. At this temperature no divergence of a relaxation
time can be observed since that time has already diverged above at the
dynamical transition. The only way to show the existence of this
transition is by estimating the configurational entropy and determining
the temperature at which that quantity vanishes. Such a task has been
successfully undertaken in the case of the random-orthogonal model
\cite{CR} which
displays a very similar behavior to the $K<0$ model and requires
the use of projection techniques, in the spirit of those introduced by
Stillinger and Weber, for the study of the potential energy landscape in
the context of glasses \cite{SW}.

\section{Conclusions}

In this paper we introduced a solvable model of a spin-glass
corresponding to the mean-field version of the nearest and next nearest
neighbor
spin-glass model. The main interest of the model is that it shows how
correlations in the couplings may completely change the character of the
spin-glass transition from continuous replica symmetry breaking to a
one-step transition. We analyzed two cases, the positive model $K>0$
with a phase transition and behavior similar to that of the SK model and
the $K<0$ model with ideal mode-coupling behavior and a phase transition
similar to the ROM or to the $p$-spin model. The different character of
the transition in both models can be ascribed to the different behavior
of the eigenvalue density distribution close to the maximum eigenvalue
or threshold. For the $K>0$ the distribution vanishes at the threshold
(i.e. $\lambda_+=4$) and the equilibrium distribution {\em condensates}
around configurations in the vicinity of the maximum eigenvector. For
the $K<0$ the eigenvalue density diverges at the threshold
(i.e. $\lambda_-=0$). In that case, the equilibrium configuration {\em
condensates} around an extremely large number of eigenvectors with
eigenvalue close to 0. Hence, the phase space splits into a very large
number (exponentially large with the size of the system) of ergodic
components or phases, these phases corresponding to different
eigenvectors with eigenvalue very close to the threshold which
extensively contribute to the configurational entropy.  Obviously,
eigenvectors of the coupling matrix  never coincide with possible
configurations of the Ising system, so this argumentation must be
taken only as a rough picture. 
Nevertheless, the idea that the type of eigenvalue
distribution determines the character of the transition seems quite
intuitive. Actually, if one considers spherical instead of Ising spins
then the transition disappears for the $K<0$ model but persists for the
$K>0$ case. For the $K<0$ model the transition disappears because there
is no longer a vanishing  of the configurational entropy  (the classical
nature of spins allows for a negative entropy). The same mechanism
occurs in the ROM where the one-step transition with Ising spins
disappears in case of spherical spins. In this case, the density
of eigenvalues is given by
$\rho(\lambda)=\frac{1}{2}\delta(\lambda-1)+\frac{1}{2}\delta(\lambda+1)$
showing that the density diverges in the vicinity of the maximum
eigenvalue. 

What is the real connection between the density of eignevalues and the
order of the transition? The mechanism which makes a continuous
transition  become first order (in the order parameter) in spin-glass
theory is well known and based on the role of the so called complexity
or (sometimes misleadingly) configurational entropy \cite{KTW,MO,ME}. In
mean-field theory the complexity $S_c(F,T)$ at a given free energy and
temperature corresponds to the logarithm of the number of TAP free
energy minima solutions with that free energy and temperature. The
complexity defines the so called potential $\Phi(F,T)=F-TS_c(F,T)$. This
quantity display two types of different behaviors at low enough
temperatures. For models with continuous replica symmetry breaking
$\Phi$ has a single minimum at the equilibrium free energy value $F_{\rm
eq}(T)$ while for models with a one-step scenario this function may
display a minimum at a threshold value $F^*(T)$ higher that the
equilibrium free energy $F_{\rm eq}(T)$. The difference
$\beta(F^*(T)-F_{\rm eq}(T))$ defines the complexity at that temperature
$S^*(T)=S_c(F^*(T),T)$ which is positive for one-step models but
vanishes in continuous models.  Because at high free energies
$\Phi(F,T)\sim F$ it is clear that the abundance of metastable solutions
makes $S_c$ always larger raising the possibility that the potential
$\Phi(F,T)$ displays a minimum above the equilibrium free
energy. Therefore, a divergence of the eigenvalue density at the lowest
value goes in the appropriate direction of generating a minimum in
$\Phi$ and hence making the transition become first-order in the
order parameter.

As for further directions of interest we mention the investigation of
the existence of a model interpolating between the $K>0$ and $K<0$
models. This would be very interesting to understand better the
mechanism which makes the spin-glass transition change from a continuous
one to a discontinuous one, as well as to connect this change with the behavior of
the density of eigenvalues close to the threshold. In this direction it
would be also interesting to investigate in a general way the type of
phase transition for a generic eigenvalue distribution using analytical
techniques such as those developed for the ROM model \cite{ROM}.

 It would also be interesting to investigate the behavior of the
different terms in the TAP expansion to see how this change of behavior
occurs and how this is related to the geometrical properties of the free
energy landscape. Note that the information contained in the density of
eigenvalues is closely related to the topological features of the energy
landscape. Therefore, it is reasonable to think that the nature of the
transition in these models arises from the geometrical properties of the
energy (and hence, the free energy) landscape.

Finally, we would like to investigate the behavior of this model on a
finite dimensional lattice. In the simplest way this can be done taking
$J$ to be  nearest-neighbor Gaussian matrix and defining $K$ according to
(\ref{KJ}). It would be very interesting to see how the two universality
classes describing the mean-field behavior are modified in the presence
of activated processes and how they manifest themselves 
in the dynamical properties
of realistic systems. This could give additional understanding
about why some models behave in one way or another and how topological
properties of the landscape influence the dynamical behavior of real
systems.

{\bf Acknowledgments}.  We are grateful to A. Lef\`evre and M. Sales for a
careful reading of the manuscript. F.R is supported by the Ministerio de
Ciencia y Tecnolog\'{\i}a in Spain, project BFM2001-3525 and Generalitat
de Catalunya.


\baselineskip =18pt

\begin{thebibliography}{0}



\bibitem{SK} D. Sherrington and S. Kirkpatrick, Phys. Rev. Lett. {\bf 35}, 1792 (1975); 
Phys. Rev. B, {\bf 17}, 4348 (1978) 

\bibitem{REVIEW} K.~Binder and A.P.~Young, Rev. Mod. Phys. {\bf 58}, 801
(1986); M.~M\'ezard, G.~Parisi and M.A.~Virasoro, {\em Spin Glass
Theory and Beyond} (World Scientific, Singapore 1987); K.H.~Fischer
and J.A.~Hertz, {\em Spin Glasses} (Cambridge University Press,
Cambridge 1991); V. Dotsenko, {\em Introduction to the Replica Theory
of Disordered Statistical Systems}, (Cambridge University Press,
Cambridge 2000)

\bibitem{EA} S. F. Edwards and P. W. Anderson, J. Phys. F {\bf 5}, 965 (1975). 

\bibitem{PARISI} G. Parisi, Phys. Rev. Lett. 43, 1754-1756 (1979); J. Phys. A {\bf 13}, 1101 (1980).

\bibitem{DROP} D.S. Fisher and D.A. Huse, Phys. Rev. Lett. {\bf 56}
 1601 (1986); A.J. Bray  and M.A. Moore, Phys. Rev. Lett.
{\bf 58}, 57 (1987)

\bibitem{WIGNER} M. L. Mehta {\em Random Matrices and the Statistical 
Theory of Energy Levels} (New York: Academic) (1967).

\bibitem{SKSPHER} J. M. Kosterlitz, D.J. Thouless and R.C. Jones 
Phys. Rev. Lett. {\bf 36} 1217 (1976). 

\bibitem{LITTLE} W. A. Little, Math. Biosci. {\bf 19}, 101 (1974).

\bibitem{RHOCOM} In the nonsymmetric case the density of eigenvalues
can be calculated within the replica framework.

\bibitem{HOP} J. J. Hopfield, Proc. Natl. Acad. Sci. {\bf 79}, 2554 (1982).

\bibitem{CMPP} S. Cabasino, E. Marinari, P. Paolucci and G. Parisi, 
J. Phys. A (Math. Gen.){\bf 21}, 4201 (1988).

\bibitem{BPR} R. Brunetti, G. Parisi and F. Ritort, Phys. Rev. B {\bf 46},
5339 (1992).  

\bibitem{AGM} J. H. Gibbs and E. A. Di Marzio,
J. Chem. Phys. {\bf 28}, 373 (1958); G. Adams and J.H. Gibbs, {\it
J. Chem. Phys.} {\bf 43}, 139 (1965).

\bibitem{K} W. Kauzmann, {\it Chem. Rev. } {\bf 43}, 219 (1948).

\bibitem{MCT} E. Leutheusser, Phys. Rev. A {\bf 29}, 2765 (1984);
T. R. Kirkpatrick, Phys. Rev A {\bf 31}, 939 (1985); W. G\"otze and
L. Sjogren, Rep. Prog. Phys. {\bf 55}, 241 (1992).

\bibitem{KTW} T.R. Kirkpatrick and P.G. Wolynes, Phys. Rev. B {\bf 36},8552 (1987);
 T.R. Kirkpatrick, D. Thirumalai and P.G. Wolynes, Phys. Rev. A {\bf
 40}, 1045 (1989).

\bibitem{ROM}  E. Marinari, G. Parisi and F. Ritort, J. Phys. A (Math. Gen.) {\bf 27},
7647 (1994).

\bibitem{POTTS} E. De Santis, G. Parisi and F. Ritort,
J. Phys. A (Math. Gen.) {\bf 28}, 3025 (1995).

\bi{Q} F. Ritort, Phys. Rev. B {\bf 55}, 14096 (1997).  

\bibitem{CHS} A. Crisanti, H. Horner and H. J. Sommers, Z. Phys. B {\bf 92},
257 (1993). 

\bibitem{OPF} M. Picco, F. Ritort and M. Sales, Eur. Phys. J. B. {\bf 19},
565 (2001).

\bibitem{CR} A. Crisanti and F. Ritort, Preprint cond-mat/0110259. 

%Europhys. Lett. {\bf 51}, 147 (2000).

\bibitem{SW} F. H. Stillinger and T. A. Weber, Phys. Rev. A {\bf 25}, 978 (1982).

\bibitem{MO} R. Monasson,  Phys. Rev. Lett. {\bf 75}, 2847 (1995).

\bibitem{ME} M. Mezard, Physica A {\bf 265}, 352 (1999). 

\end{thebibliography}
\end{document}